\documentclass[aps,pra,reprint,twocolumn,amsfonts,amssymb,amsmath,floatfix,tightenlines]{revtex4-1}
\usepackage{graphicx}
\usepackage{pdfpages}
\usepackage{relsize}
\usepackage{bm}
\usepackage[colorlinks=true,citecolor=blue,urlcolor=blue]{hyperref}

\makeatletter
\AtBeginDocument{\let\LS@rot\@undefined}
\makeatother

\numberwithin{equation}{section}

\makeatletter
\AtBeginDocument{\let\LS@rot\@undefined}
\makeatother

\begin{document}
\title
{Quantum Phase Transitions in Proximitized Josephson Junctions}

\author{Chien-Te Wu$^{1,2}$}
\author{F. Setiawan$^{1}$}
\author{Brandon M. Anderson$^{1}$}
\author{Wei-Han Hsiao$^{1}$}
\author{K. Levin$^{1}$}
\affiliation{$^1$James Franck Institute, University of Chicago, Chicago, Illinois 60637, USA}
\affiliation{$^2$Department of Electrophysics, National Chiao Tung University, Hsinchu 30010, Taiwan, Republic of China}

\begin{abstract}

We study fermion-parity-changing quantum phase transitions (QPTs) in
platform Josephson junctions. 
These QPTs, associated with zero-energy bound states,
are rather widely observed experimentally. They emerge 
from numerical calculations frequently without detailed microscopic
insight. Importantly, they may incorrectly lend support to
claims for the observations of Majorana zero modes.
In this paper we present
a fully consistent solution of the
Bogoliubov-de Gennes equations for a multi-component
Josephson junction. This provides
insights into
the origin
of the QPTs.
It also makes it possible to assess the standard
self energy approximations which are widely used to
understand proximity coupling in topological
systems.
The junctions we consider 
are complex and chosen to
mirror 
experiments.
Our full proximity calculations associate the mechanism behind the QPT as
deriving from a spatially extended, proximity-induced magnetic ``defect".
This defect arises because of the insulating
region which effects a local reorganization of the bulk magnetization
in the proximitized superconductor.
Our results suggest more generally that
QPTs
in Josephson junctions
generally do not require the existence of spin-orbit coupling
and should not be confused with, nor are they indicators of, Majorana physics.
\end{abstract}


\maketitle

\section{Introduction}

Josephson junction geometries, particularly in the presence
of magnetic fields, are becoming of greater interest in the 
search for and confirmation
of topological superconductors.
Often present in these spinful Josephson junctions are
fermion-parity switches.
A fermion-parity switch is a quantum phase transition
(QPT) where the superconducting condensate can lower
its ground state energy by incorporating an unpaired
electron and changing the number of electrons in the ground
state from even to odd. 
This results in zero-energy bound states and energy level crossings which are 
protected
(due to fermion parity) and
are, thus, not lifted by a superconducting gap.
The QPTs of interest here are claimed
by some \cite{Sau-Demler,Spanish,Lieber2017} to be important indicators of topological phases. Others 
\cite{Glazman2015,Marcus} argue that they  
may be more ``accidental" and they, rather,
make it difficult to distinguish the interesting Majorana 
quasi-particles from
conventional fermionic subgap states.

\begin{figure*}
\includegraphics[width=6.5in,clip]
{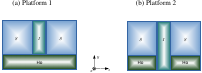}
\caption{Schematic illustration of proximitized Josephson junctions.
Two types of platforms are considered in this work.
Platform 1 is a typical Josephson junction with 
an insulating barrier placed on top of 
a conical ferromagnet (Ho) substrate with effective spin-orbit and Zeeman
coupling. 
In platform 2, two conical-ferromagnet/superconductor
bilayers are separated by an insulating barrier.
Both platforms are assumed to be finite 
along the $y$ and $z$ directions, but infinite along the $x$-direction.}
\label{fig:1a}
\end{figure*}

Because it is not clear the extent to which these QPTs 
relate to topological superconductivity, and to clarify their origin more
generally, in the present paper we investigate their behavior in Josephson
junctions. 
We study 
a ``proximitized" Josephson junction which includes two
host superconductors
which induce pairing in a
substrate medium. The latter contains both Zeeman and spin-orbit
coupling (SOC), as necessary for topological phases.
We solve the
the full set of BdG equations in this multi-component
system. This makes it possible
to assess the standard
self energy approximations 
\cite{Saurobustness,Stanescu2,Stanescu3,Stanescu_2011,Hell}
which are widely used to
implement proximity coupling in topological materials.

Our fully self consistent treatment allows us to compute the
induced magnetization 
$\bf{m}(\bf{r})$, 
in the junction. These calculations indicate that the non-topological zero energy
bound states
are confined to regions where 
$\bf{m}(\bf{r})$,
is most inhomogeneous.
This suggests a scenario for the bound state origin
involving a proximity-induced
``magnetic" 
defect. 
Although we build on some of the formal similarities, it
is important to contrast this with
an inserted magnetic impurity in a superconducting host \cite{yu1965bound,shiba1968classical,Sakurai}.
Here we associate the ``magnetic defect" with the insulating region which effects
a local reorganization of the magnetization in the proximitized medium.
Central to obtaining quantum phase transitions in this picture is the presence of Zeeman fields in the junctions.

We consider two types of geometries as shown in Fig.~\ref{fig:1a}.
The first platform contains a holmium (Ho) substrate
in the $xz$ plane on top of which are placed
two superconductors 
separated by an insulator I.
Although we believe our results to be quite general,
for definiteness we take the proximitized
superconductor
to be
a conical magnet
(Ho) which the spintronics community
has established \cite{Robinson}
exhibits well-controlled and
well-characterized proximity coupling.
Importantly, in Ho the Zeeman and SOC are intrinsically
present. The conical magnetism implies that the SOC is effectively
one dimensional (1D), as distinguished from Rashba SOC.
And interestingly it is possible to control the 1D SOC with a very
benign or non-intrusive ``knob" \cite{Robinson2},
through the unwinding of the conical
order.

For the second platform, we
consider two Ho-Superconductor (Ho-S) bilayers
in contact with an intermediate
insulating layer.
Both platforms are assumed to have
finite thicknesses in the $y$ and $z$ directions as shown in
Fig.~\ref{fig:1a} and are
taken to be infinite along $x$.
We stress that, in both platforms,
two conventional
superconductors
are used to induce superconducting order in Ho
which has no intrinsic pairing.
This proximity-induced superconductivity is
characterized by solving the Bogoliubov-de Gennes (BdG) equations
with the incorporation of full self-consistency.
Introducing two different junction configurations allows us
to further contrast the behavior of trivial versus topological
zero energy bound states. 
In both platforms and in topological phases one can access
bound states which are associated with Majorana particles. We find that
platform 1 also hosts trivial zero energy bound states which we associate
with magnetization inhomogeneities. Platform 2, by contrast, does not
contain these trivial crossings; rather it only
hosts true (albeit, hybridized) Majorana states.

Our QPTs
should also be relevant to the broader, and topical
issue of
zero bias conductance peaks. Here, too, there are reports of QPT,
often not associated with topological phases \cite{LiuSauref1,LiuSauref2,LiuSauref3}. 
While the literature on topological Josephson junctions has focused on
proximity-induced superconductivity primarily in nanowires, in order
to emphasize QPTs and level crossings, the junctions
we contemplate extend indefinitely into the $x$ dimension; 
this conveniently introduces a variable $k_x$ thereby providing
continuous tuneability and, importantly, wider access to zero-energy
bound states, fermion-parity switches and level crossings.

Notably, we find that all this interesting physics 
arises
via
proximity-induced superconductivity. 
While
the nature and location of the zero-energy bound states were not obvious a 
priori, one
might have erroneously anticipated 
that they relate to states within
the insulator. However,  
we find them here to be almost exclusively localized in the 
proximitized superconductors (in this case, Ho).

\subsection{Background Literature}

Relevant to the work in this paper is our earlier
study \cite{ourPRB} of a two component Ho-S proximity system. There
we have shown
that
the end result is a heterostructural nodal topological
superconductor.
By studying the fully self consistent BdG equations in finite size systems, we 
have
demonstrated how excitation gaps and general features of topological energy dispersion
along with Majorana zero modes are found to be present.

Turning now to Josephson junctions, 
fermion parity shifts in QPTs
and associated energy level crossings
have appeared most commonly in the literature
in two related (non-topological) contexts 
associated with 
localized magnetic impurities (such as the Shiba state) as well as 
in tunnel junctions involving quantum dots (QD) 
\cite{Glazman2,Glazman2015,Marcus,Lieber2013,Lieber2017,QuantumDotDeacon}
with strong
Coulomb correlation. In the latter context the quantum dots are thought to contain
trapped spin 1/2 single electrons which play a similar role
as magnetic impurities.
They, thus, can host
two distinct ground states \cite{yu1965bound,shiba1968classical,Sakurai}, which in turn can lead
to fermion parity shifts.

In a somewhat different vein, based on the physics of Shiba states, Sau and Demler \cite{Sau-Demler} 
suggest that non-magnetic
impurities
may be used as a probe of topological superconductivity 
\cite{Majorana1,Majorana2,Majorana3}.
They argue that in the topological phase, a non-magnetic impurity
will lead to distinctive pairbreaking due to the associated $p$-wave
symmetry.  
Yazdani and co-workers \cite{Yazdanigroup1,Yazdanigroup2} have inverted this situation in a sense by
using magnetic impurities to bind fermions into a 1D
Kitaev-chain which can then play the role of a nanowire with
topological superconducting order.

Parity switches and related zero-energy bound states, of interest
in the present paper, have also
led to a lively debate about recent experiments 
\cite{Majoranaexpts1,Majoranaexpts2} which claim evidence
for Majorana fermions. These have mainly focused on zero bias conductance 
peaks.
Liu and co-workers have recently \cite{LiuSau} studied how Andreev
bound states associated with quantum dots may produce near-zero-energy midgap states
as the Zeeman splitting and/or
chemical potential are tuned. They find the behavior as a function of magnetic
field and chemical potential is sufficiently complex so that one cannot arrive
at simple governing equations. These zero-energy Andreev bound states (ABSs) mostly
appear in the nontopological regime;
here the quantum dot was assumed to have no Coulomb blockade behavior.

In the trivial phase, others \cite{LiuSauref1,LiuSauref2,LiuSauref3} have demonstrated how near-zero-energy states may arise
in a spin-orbit coupled nanowire (in the presence of a magnetic field)
and associated these with disorder effects or details in the wire's end or even temperature.
Because of these and related papers, it is natural for there to
be concern that
zero-energy bound states related to parity switches
can give rise \cite{Glazman2015} to
features which could be confused with topological phases;
thus, they need to be well understood
before they can be safely disregarded.

\subsection{Proximity Effects}

We stress that most of the current thinking about topological superconductivity
is based on the proximity effect. 
A fully complete and
detailed treatment
of this proximitization is
complicated \cite{Klapwijk}. Moreover, for the case of
Josephson junctions we know of no prior, fully precise calculations in the
topological literature.
Proximity
effects are conventionally handled \cite{Saurobustness,Stanescu2,Stanescu3,Stanescu_2011,Hell}
through a simplification, by integrating out the (host)
superconducting degrees of freedom. This introduces 
an effective
self energy term in the proximitized medium. In this way a pairing gap
is assumed to be present, but it is generally taken to be piece-wise constant (or
zero) in different
regions of the heterostructure.

How good are these approximations and how accurately do they represent the
more exact physics? These are questions we address in this paper
in the context of Josephson junctions.
Here we use an
alternate methodology
\cite{halterman2001} developed for ferromagnet-superconductor
junctions in spintronics. This involves
a complete solution of the BdG equations
for a given heterostructure. As is physical,
the pairing attraction is assumed present only
in the host superconductors.
Proximitization introduces a non-vanishing value for
the so-called pair amplitude $F({\bf r}) = \langle\psi_{\downarrow}({\bf r})\psi_{\uparrow}({\bf r})\rangle$.
Importantly, these induced pairing correlations which appear
in a non-superconducting
system can be
sufficiently strong so as to produce a Meissner effect~\cite{Klapwijk}.
Unlike in an intrinsic superconductor, however, the phase coherence of
these pairing correlations is maintained only over a restricted length scale
~\cite{halterman2002}.

To understand the proximity findings in the present paper
more deeply, we present
a comparison of the full BdG solution with the widely used self-energy
scheme
\cite{Stanescu1,Stanescu2,Stanescu3,Hell}.
This leads to
a 3 component SIS structure (with infinite extent in the $x$ direction).
What emerges from this comparison of the full versus the
approximate treatment of proximity effects is that for 
both platforms there are broad-based similarities. For the
first the QPT we find are associated with the trivial, 
while for the second they appear in the topological
phase. We note that the energy dispersions in the two theoretical approaches
appear rather differently and
it is often difficult to access the same parameter regime, particularly
in the second platform.
This is discussed in more detail in
Section IV.

\subsection{Zero-Energy crossing as a zero dimensional topological phase
transition}\label{sec:ZEC}

It is important to understand the zero-energy crossings (ZECs)
in these SIS junctions
in more depth. It is known that such crossings in the energy
spectrum
are signatures
of a (zero dimensional) topological phase transition involving a change in
fermion parity.

As in a topological superconductor, a fermion-parity changing
QPT can
be understood through a change in the Pfaffian of the BdG Hamiltonian.
In particular, near a crossing point
$k_{\mathrm{c}}$
we may project the Hamiltonian onto the subspace spanned
by the corresponding two crossed states denoted
$|\pm\rangle$
so that the
resulting two-band effective Hamiltonian is given by
\begin{align}
H^{mn}_{\mathrm{eff}} = \langle m|\nabla_{k_x}H|n\rangle \delta k_x,
\end{align}
where $m,n = \pm$. By diagonalizing the effective Hamiltonian, we obtain
\begin{align}\label{eq:Heff}
H_{\mathrm{eff}} = A \delta k_x \sigma_z,
\end{align}
where $\sigma_{x,y,z}$ are the Pauli matrices acting in the $|\pm\rangle$ subspace. By performing a basis rotation, we can write Eq.~\eqref{eq:Heff} into its skew-symmetric form:
\begin{equation}
H_{\mathrm{eff}} = A \delta k_x \sigma_y.
\end{equation}
The Pfaffian of this skew-symmetric Hamiltonian is simply sgn($A$).
Since the
sign of
$H_{\mathrm{eff}}$
changes at the crossing point (unless $A=0$), its Pfaffian changes.
Moreover, since
$H_{\mathrm{eff}}$
is a $2 \times 2$ block in the full Hamiltonian,
the Pfaffian of the full Hamiltonian changes as well. Hence,
this is a topological phase transition between two different
``phases" in class BDI.

\begin{figure*}
\includegraphics[width=4.5in,clip]
{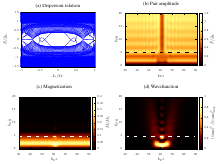}
\caption{Results from the first Josephson platform
for (a) the energy dispersion versus $k_x$, (b)
the pair amplitude, (c) the magnetization and, (d)
the wavefunction at a crossing point.
The phase difference between the two superconducting
leads is set to be zero.
In panel (a) the flat bands
on two wings of this plot are associated with Majorana edge modes
whereas the linear crossings in the central region
are associated with the localized ``impurity" Andreev bound states,
caused by the insulating defect. 
In panel (c) there is a local maximum in the magnetization in the Ho region just below the insulator.
We normalize the magnetization $M$ by $M_0\equiv \mu_B (n_{\uparrow}-n_{\downarrow})/(n_{\uparrow}+n_{\downarrow})$,
where $n_\sigma$ is the spin density.
The bottom right panel shows a typical localized wavefunction
associated with a crossing state at $k_x = 0.36 k_F$; this is in
the same region as the magnetization inhomogeneity.
The boundary between the Josephson junction and the substrate
is represented by dashed lines.
We choose the thickness of the two superconductors
along the $y$-axis to be $d_S = 75k_F^{-1}$
and the length of them to be $l_S=60k_F^{-1}$ along the $z$-axis.
The superconducting coherence length is $\xi = 2/(\pi\Delta_0)= 10k_F^{-1}$.
The chemical potential
in the insulator
is chosen to be $\mu_I = - 2E_F$ and the exchange
interaction in Ho is $0.2E_F$. 
The insulating
region has a width $\ell_I = 0.1\xi$ (from $z = 60k_F^{-1}$ to $z = 61k_F^{-1}$).
The thickness of the Ho substrate is $d_{\rm Ho} = 0.5 \xi$ (from $y = 0$ to $y = 5k_F^{-1}$).
}
\label{fig:2}
\end{figure*}

\begin{figure*}
\includegraphics[width=4.5in,clip]
{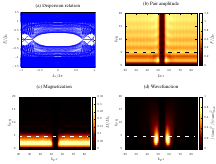}
\caption{
Results for the second Josephson platform in which the zero energy
crossing corresponds to two Majorana states in the middle of the Junction.
(a) Energy dispersion versus $k_x$,
(b) The pair amplitude, (c) magnetization and
(d) the wavefunction at a crossing point.
The phase difference between two superconducting
leads is set to be zero.
In panel (a) the flat bands
on two wings of this plot are associated with Majorana edge modes
at the ends of Holmium away from the junction.
whereas the linear crossings are associated with 
two Majorana modes at the Ho/I interfaces.
The self-consistent pair amplitudes,
shown in panel (b)
are suppressed in the insulating region.
The magnetization (c) turns on abruptly just
beyond the insulating region. Panel (d) shows that the
localized wavefunction amplitude ($k_x = 0.90k_F$) becomes large just where
the magnetic inhomogeneity appears.
The boundary between the S layers and the Ho layers
is represented by dashed lines.
We choose the thickness of the two superconductors along the $y$-axis to be $ d_S = 75k_F^{-1}= 7.5\xi$.
where $\xi$ is the superconducting coherence length. The thickness of the Holmium is $d_{\rm Ho} = 5 k_F^{-1} = 0.5\xi$. The length of the Holmium/superconductor bilayers along the $z$-axis is taken to be $\ell_S = \ell_{\mathrm{Ho}} = 40 k_F^{-1} = 4 \xi$. The chemical potential of the superconductor and Holmium are taken to be $\mu_{S} = \mu_{\mathrm{Ho}} = E_F$. 
The chemical potential
in the insulator, which separates two bilayers,
is chosen to be $\mu_I = -2E_F$ and its width is $\ell_I = 5k_F^{-1}$.
The exchange field in the conical ferromagnets is $0.2E_F$.
}
\label{fig:3}
\end{figure*}

\subsection{Self Consistency in Bogoliubov-de Gennes Theory}

In this overview section we outline the general structure of
self-consistent BdG schemes.
More specific details
of this approach are given in the following section.
Self-consistency
involves
solving for the pair amplitude  $F({\bf r}) \equiv \langle\psi_{\uparrow}(\bf{r})\psi_{\downarrow}({\bf r})\rangle$ at a microscopic level
in terms of the attractive interaction.
This is to be contrasted with the more standard BdG approach in the
recent literature (on topological superconductivity) which is
to solve the same equations for the wavefunctions and
energies, but not include the feedback of these
wavefunctions into the determination of the gap.

We stress that imposing a self-consistent gap equation is fundamental
to BCS theory and its extensions to non-uniform situations such as
in Gor'kov theory or the equivalent BdG approach.
Proximitized superconductors necessarily involve spatially dependent gap
functions; fixing the gap value to be constant in a certain region
of the sample may make the calculations easier, but they may miss
essential physics which is particularly relevant
at interfaces and boundaries. Here we use these
(sometimes abrupt) spatial variations in the pair amplitude and in derived
quantities such as the magnetization to establish correlations
and thereby provide insight into quantum
phase transitions.

Our system can be described by the following mean-field Hamiltonian 
\begin{align} 
\mathcal{H} &= \int d^3 \mathbf{r} \psi_{\sigma}^{\dagger}(\mathbf{r}) \left( H_0 + H_{\rm Zeeman} \right)_{\sigma\sigma^\prime} \psi_{\sigma^\prime}(\mathbf{r}) \nonumber\\
 	    &+\left[\Delta(\mathbf{r})  \psi_{\uparrow}^{\dagger}(\mathbf{r}) \psi_{\downarrow}^{\dagger}(\mathbf{r}) +\textrm{H.c.}\right],\label{eq:full}
\end{align}
where $\psi_{\sigma}^\dagger(\mathbf{r})$ ($\psi_{\sigma}(\mathbf{r})$) with spin $\sigma=\uparrow,\downarrow$ are fermionic creation (annihilation) operators and $H_0$ is the single particle 
contribution.  
An attractive pairing interaction is only present in
the superconductors.
Here $H_{\rm Zeeman} = {\bf h} \cdot {\bm \sigma}$ is the Zeeman term in the full Hamiltonian, Eq.~(\ref{eq:full}), where
${\bm \sigma}=\left(\sigma_x,\sigma_y,\sigma_z\right)$ are Pauli matrices
and ${\bf h}$ is the exchange interaction associated with a conical
magnet 
which is given by
\begin{equation}
\label{exchange}
{\bf h}=h_0\left\{\cos\alpha\hat{z}+\sin\alpha\left[\sin\left(\frac{\beta z}{a}\right)\hat{x}+\cos\left(\frac{\beta z}{a}\right)\hat{y}\right]\right\},
\end{equation}
Here $\beta$ is the periodicity of the helix and $a$ is the lattice constant along the $c$ axis of Holmium. As in the literature~\cite{chiodi,robinson2009}, the opening angle $\alpha=\pi/2$.
From our previous analysis~\cite{ourPRB}, this puts our model Hamiltonian in the BDI class. The spiral magnetic order introduces a combination of one-dimensional spin-orbit coupling (SOC) and Zeeman field~\cite{NadjPergProposal2013,IvarMajorana2012}. This can be shown by applying a gauge transformation $\psi_\uparrow \rightarrow e^{-i\beta z/2a} \psi_{\uparrow}$ and $\psi_\downarrow \rightarrow e^{i\beta z/2a} \psi_{\downarrow}$ which transforms the single-particle Hamiltonian into
\begin{equation}
H'_{0} + H'_{\mathrm{Zeeman}} = H_{0} - v_{so}\sigma_z i \partial_z + m v^2_{\mathrm{so}}/2 + \mathbf{h}'\cdot \bm{\sigma},
\end{equation}
where the effective Zeeman strength is $\mathbf{h}' = h_0 (\sin\alpha \hat{\bf{x}} + \cos \alpha \hat{\bf{z}})$ and the spin-orbit-coupling strength is $v_{\mathrm{so}} = \frac{\beta}{2ma}$. In this way, the magnetic spiral period $\lambda = 2 \pi a/\beta$ sets the strength of the spin-orbit coupling  $v_{\mathrm{so}} = \pi/(m \lambda)$.

The
BdG equation in matrix form is given by,
\begin{align}
&\left(\begin{array}{cccc}
H_{0} & h_{x}-ih_{y} & 0 & \Delta\\
h_{x}+ih_{y} & H_{0} & -\Delta & 0\\
0 & -\Delta^{\ast} & -H_{0} & -h_{x}+ih_{y}\\
\Delta^{\ast} & 0 & -h_{x}-ih_{y} & -H_{0}
\end{array}\right)
\left(\begin{array}{c}
u_{n\uparrow} \\
u_{n\downarrow} \\
v_{n\uparrow} \\
v_{n\downarrow} \\
\end{array}\right)\nonumber \\
&=E_n
\left(\begin{array}{c}
u_{n\uparrow} \\
u_{n\downarrow} \\
v_{n\uparrow} \\
v_{n\downarrow} \\
\end{array}\right),
\label{matrixBdG}
\end{align}
where $u_{n\sigma}$ ($v_{n\sigma}$)
are quasi-particle (hole) wavefunctions.
We have suppressed the position label, $\bf r$, in
Eq.~(\ref{matrixBdG}). The Hamiltonian in Eq.~\eqref{matrixBdG} can be written more compactly in the Nambu basis as 
\begin{equation}
\mathcal{H_{\mathrm{BdG}}} = H_0 \tau_z + \mathbf{h}.\bm{\sigma} \tau_z +[i\Delta \sigma_y\tau_+ + \mathrm{H.c.}], 
\end{equation} 
where $\bm{\sigma}$ and $\bm{\tau}$ are the Pauli matrices acting in the spin and particle-hole subspaces, respectively.

Now the essence of a self-consistent BdG approach is to obtain
the superconducting pair potential (or ``gap" parameter), $\Delta({\bf r})$,
microscopically from the attractive interactions:
\begin{equation}
\Delta ({\bf r})  \equiv  g({\bf r}) F({\bf r}),
\end{equation}
where  $g({\bf r})$ is the coupling constant which
vanishes outside the superconductor~\cite{halterman2001}
and 
\begin{equation}\label{eq:Deltaself}
F({\bf r}) =  \sum_{\epsilon_n<\omega_D} \left[u_{n\uparrow}({\bf r})v_{n\downarrow}^*({\bf r})-u_{n\downarrow}({\bf r})v_{n\uparrow}^*({\bf r})\right] \tanh\left(\frac{\epsilon_n}{2T}\right)
\end{equation}
is the pair amplitude.
Note that the Debye frequency $\omega_D$ is the energy cutoff and $T$ is the temperature (we set $\omega_D=0.1E_F$ and $T=0$ in this paper).
Important for the present purposes is that in our proximitized
superconductors, the pair amplitude
is non-zero, even though there is a vanishing order parameter.

Another important physical property which involves $u_{\sigma}$ and $v_{\sigma}$
is the position-dependent magnetization.
\begin{subequations}\label{eq:magnetization}
\begin{align}
m_x({\bf r}) =& - \mu_B \sum_{n}\left(v_{n\uparrow}({\bf r})
v_{n\downarrow}^{\ast}({\bf r})+v_{n\downarrow}({\bf r})
v_{n\uparrow}^{\ast}({\bf r})\right),\\
m_y({\bf r}) =& i \mu_B 
\sum_{n}\left(v_{n\uparrow}({\bf r})
v_{n\downarrow}^{\ast}({\bf r})-v_{n\downarrow}({\bf r})
v_{n\uparrow}^{\ast}({\bf r})\right),\\
m_z({\bf r}) =& - \mu_B \sum_{n}\left(|v_{n\uparrow}({\bf r})|^2-
|v_{n\downarrow}({\bf r})|^2\right),\\\nonumber
\end{align}
\end{subequations}
where $\mu_B$ is the Bohr magneton.
This is a central quantity in the present paper.
The spatial dependence of the magnetization $ M({\bf r})= \sqrt {m_x^2 + m_y^2 +m_z^2}$
is associated with
a magnetic screening cloud in the superconductor. 
Of interest is how in a proximitized superconductor, the magnetization
can be reorganized (say, by insulating barriers, defects and
interfaces) in a way which
is readily quantified.

\section{BdG Approach for proximity calculations}
\subsection{Numerical procedure}

The single particle term in the full Hamiltonian which appears in
Eq.~(\ref{eq:full}) is 
\begin{equation}
H_0 = -\frac{\nabla^2}{2m} - \mu(y,z).
\end{equation}
This describes free fermions of mass $m$. Throughout the paper we adopt the natural units $\hbar=k_B=1$.
The insulator is associated with the position-dependent quantity $U_0(y,z)$
which reflects a localized shift in the chemical potential. We use a single positive Fermi energy $E_F$
for the chemical potentials of the superconducting layers $\mu_S$ and the substrate $\mu_{\mathrm{Ho}}$.
The chemical potential $\mu_I$ for the insulator should be negative.
More precisely, the shift of the insulating Fermi level takes the form $U_0(y,z)=$
\begin{equation}
E_F-\left(E_F-\mu_I\right)\Theta\left(y-d_{\rm Ho}\right)\Theta\left(z-\ell_{S}\right)\Theta\left(\ell_{S}+\ell_I-z\right)
\end{equation}
in the first platform and
\begin{equation}
E_F-\left(E_F-\mu_I\right)\Theta\left(z-\ell_{S}\right)\Theta\left(\ell_{S}+\ell_I-z\right)
\end{equation}
in the second platform.
We define $d_{\mathrm{Ho}}$, $d_{S}$, $\ell_{I}$, and $\ell_{S}$ as respectively the thickness 
of the Ho substrate (along the $y$-direction), thickness of the superconductor (along the $y$-direction), the length of the insulator (along the $z$-direction), and the length of the superconductor (along the $z$-direction).
(For simplicity, we take the two S layers in these platforms
to be the same along both $y$- and $z$-direction).
The origin of the $y-z$ coordinate system is at the bottom left corner for
both junction configurations as shown in Fig.~\ref{fig:1a}.

Because the Hamiltonian is translationally invariant along $x,$ the
proposed wavefunction in the $x$ direction is in the form of $e^{ik_{x}x}.$ Therefore,
\begin{equation}
H_{0}=-\frac{1}{2m}\left(\partial_{y}^{2}+\partial_{z}^{2}\right)+\frac{k_{x}^{2}}{2m}- \mu(y,z).
\end{equation}
The attractive pairing interaction 
$g$ is also a function of $y$ and $z$ and
taken to be a constant associated with a bulk superconductor in the S regions.

We numerically solve the BdG eigenvalue problem following the scheme
developed in Refs.~\cite{halterman2,ourPRB,halterman2001}.
For definiteness,  we set the smallest length scale to be
in the order of $k_F^{-1}$.
We then expand both the matrix elements and
the eigenfunctions in terms of a Fourier basis.

For the quasi-particle and quasi-hole wavefunctions, we have
\begin{subequations}
\begin{align}
\tilde{u}_{n\sigma k_x}(y,z)&=&\frac{2}{\sqrt{d\ell}}\sum_{p,q}u_{n\sigma k_x}^{pq}\sin\left(\frac{p\pi y}{d}\right)\sin\left(\frac{q\pi z}{\ell}\right),\\
\tilde{v}_{n\sigma k_x}(y,z)&=&\frac{2}{\sqrt{d\ell}}\sum_{p,q}v_{n\sigma k_x}^{pq}\sin\left(\frac{p\pi y}{d}\right)\sin\left(\frac{q\pi z}{\ell}\right).
\end{align}
\end{subequations}
Note that $\tilde{u}_{n\sigma k_x}$ and $\tilde{v}_{n\sigma k_x}$ are related to 
${u}_{n\sigma}$ and ${v}_{n\sigma}$  by the relations $u_{n\sigma}=e^{ik_xx}\tilde{u}_{n\sigma k_x}$ and $u_{v\sigma}=e^{ik_xx}\tilde{v}_{n\sigma k_x}$.
General matrix elements are similarly expanded in terms of the same Fourier series.
For example, we define the matrix elements of an operator $O$ to be
\begin{eqnarray}
O^{pqp'q'} & \equiv & \langle pq|O|p'q'\rangle \nonumber\\
& =&
\frac{4}{d\ell}\int_{0}^{d}\int_{0}^{\ell}dydz \sin\left(\frac{p\pi y}{d}\right)\sin\left(\frac{q\pi z}{\ell}\right)\nonumber
\\
& &
 \qquad \times O \sin\left(\frac{p'\pi y}{d}\right)\sin\left(\frac{q'\pi z}{\ell}\right).
\end{eqnarray}

All terms in the Hamiltonian can then be expanded in this basis set.
We then have successfully transformed a set of differential equations into an algebraic matrix eigenvalue 
problem. To consider a phase difference $\phi$ between the two superconductors, our initial 
ansatz for the pair potential is: $\Delta(y,z)=$
\begin{equation}
\Delta_0\Theta\left(y-d_{\rm Ho}\right)\left[\Theta\left(\ell_S-z\right)+\Theta\left(z-\ell_S-\ell_I\right)e^{i\phi}\right],
\end{equation}
for both platforms, where $\Delta_0$ is the bulk superconducting pair amplitude.
As in our previous work~\cite{ourPRB}, we look for the self-consistent solution of $\Delta(y,z)$ iteratively.

\section{Numerical Results: Full Proximity Treatment of Josephson Junctions}\label{sec:proximity}

In this section we present the results of a full BdG
solution for the proximity junctions shown in Fig.~\ref{fig:1a}(a).
The results of the first and second platforms are shown in Figs.~\ref{fig:2} and~\ref{fig:3}, respectively, where we
assume that the two superconductors have the same phase. Figure~\ref{fig:2}(a) presents the
energy dispersion as a function of $k_x$. The darkest region ($E>\Delta_0\approx0.06E_F$)
of this panel corresponds to bulk states of the junctions, while 
the lighter region ($E<\Delta_0$) corresponds to states of the Ho substrate.
This clearly reflects that the Ho substrate,
is proximitized as the excitation gap is necessarily
smaller there than in the two superconducting leads.
This figure is in the
topological phase, (although we find similar results
in the trivial phase as well). This can be verified
from the presence of flat bands at the right
and left edges. Of great interest are the two crossings corresponding to
the QPTs, shown in the middle of Fig.~\ref{fig:2}(a).

Figure \ref{fig:2}(b) presents a color contour plot of the self-consistently 
determined pair amplitude profile
corresponding to this first platform junction configuration.
One sees that there is a non-vanishing amplitude inside Ho nearest the
superconductors [below the dashed line in Fig.~\ref{fig:2}(b)] and that 
a small pair amplitude
component penetrates into the insulating region, as well.
Below the insulating region (in Ho), the
amplitude is suppressed. Presumably this follows because the presence of
the insulating barrier interrupts proximity coupling.
We can see that this gap depression is
reflected in the magnetization plotted in Fig.~\ref{fig:2}(c). 
If we plot the wavefunctions associated with the QPT (crossing
points) as in
Fig.~\ref{fig:2}(d),
we find that they are rather well localized to the region of
inhomogeneous magnetization.

We turn now to the results of the second junction configuration as shown in Fig.~\ref{fig:3}.
This is also in the topological phase as can be seen from the presence of flat
bands in the dispersion plotted in Fig.~\ref{fig:3}(a). 
In contrast to platform 1, the zero crossings here are associated with
Majorana modes. These crossings arise from Majorana oscillations due to two hybridized Majoranas in the middle of the junction~\cite{ChengSplitting2009,SarmaSplitting2012}. While they represent different physics from the zero energy
crossings discussed previously, we provide a rather similar analysis as the
comparison between trivial (platform 1) and topological (platform 2) zero-energy crossing behavior should
be useful.

\begin{figure}
\includegraphics[width=3.5in,clip]
{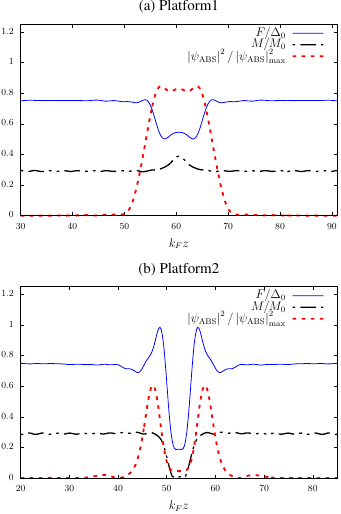}
\caption{Plots for the pair amplitude, the magnetization,
and the amplitude of $E=0$ bound state wave function as a function of position
$z$ corresponding to the
first and second platforms and with the parameters chosen from Figs.~\ref{fig:2} and~\ref{fig:3},
respectively. Here we look at fixed $k_F y = 2$ which is in the Ho layer. The bound state
in platform 1 is non-topological while that in platform 2 can be shown to be
associated with Majorana zero modes.}
\label{fig:4d}
\end{figure}

\begin{figure*}
\includegraphics[width=6.4in,clip]
{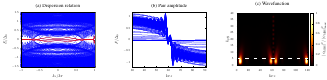}
\caption{Results for the first platform when $\phi=\pi$.
Other parameters are the same as in Fig.~\ref{fig:2}.
The left panel shows the dispersion relation. 
The two Majorana edge modes are present and now
become degenerate with two additional ``Andreev" modes
(as shown by the red lines near the zero energy).
[These were visible in  
Fig.~\ref{fig:2}(a)
where they were not
quite degenerate.] The self-consistent pair amplitudes,
plotted in the central panel,
show a clear sign change in the insulator.
The horizontal axis corresponds to the $z$-direction
and each curve in this plot is a line cut along $y$-coordinates.
In the right panel, we plot the wavefunction associated with
$k_x = 0.93 k_F$ where there are two Majorana edge modes and two Andreev modes leading
to the $4 \pi$ Josephson effect. The boundary between the S layers and the Ho layers
is represented by dashed lines.
}
\label{fig:4a}
\end{figure*}

\begin{figure*}
\includegraphics[width=6.4in,clip]
{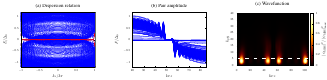}
\caption{Results for the second platform when $\phi=\pi$.
Other parameters are the same as in Fig.~\ref{fig:3}.
The figure shows, from left to right:
(a) the dispersion relation, (b) the pair amplitudes,
and (c) the wavefunction.
As in the first platform,
note the sign change in the pair amplitude as shown in panel (b). Panel (c) shows the wavefunction at $k_x = 0.96k_F$ corresponding to one of the four-fold degenerate Majorana edge modes [red lines in panel (a)].
}
\label{fig:4b}
\end{figure*}

The pair amplitude profile is shown as a color contour plot in 
Fig.~\ref{fig:3}(b). Here
the insulating component has an essentially negligible pair amplitude.
Similarly one sees from Fig.~\ref{fig:3}(c) that the magnetization is
fully absent over the insulator, but once inside the Ho it turns on
rather abruptly. The region of magnetization inhomogeneity corresponds to
the vicinity of the Ho-I boundary on either side. If we then plot the
amplitude of the crossing wavefunction as in Fig.~\ref{fig:3}(d), we
see that it is essentially localized to this inhomogeneous region,
with very small penetration into the superconducting leads.

Figure~\ref{fig:4d} presents line cuts in the Ho region of the contours
shown in the different panels (b,c,d) of the previous
two figures. Here the plots are for a fixed $y$ position in the junction.
The upper and lower panels correspond to the first and second
platforms, respectively.
All curves are for dimensionless units; by overlaying them in
this fashion one can see more clearly how the
location of the zero-energy bound states is 
in detail correlated with the inhomogeneities in 
$\bf{m}(\bf{r})$ and
$F({\bf r})$. Additionally, one can more directly compare the
behavior in the first and second junction configurations.
This enables a comparison between trivial and topological zero energy
crossings.

In the upper panel, one sees that as the insulator is approached, the
pair amplitude decreases while concomitantly the size of the
magnetization increases in precisely the same
spatial region. 
We have observed (not shown here) the expected inverse relationship between the
size of the oscillations in 
the $x$ and $y$ components of 
$\bf{m}(\bf{r})$ and the size of the pair amplitude
$F({\bf r})$.
The zero-energy bound state is, as emphasized
above, confined to the region of inhomogeneity.

The situation for the second platform (shown in the lower
panel) is more complex. Note that the
insulator is nearly ``inert",that is, it has a very small pair amplitude and very
weak
oscillating magnetism. Hence, by contrast with the upper panel,
the entire central region of the plots is nearly zeroed out.
As a consequence the
bound state (hybridized Majorana) wavefunction is restricted to a region of
width comparable to that in the first platform, but with
the central region replaced by a ``hole" across the extent of
the insulator.

\subsection{Topological Josephson junctions}

We turn now to junctions with a $\pi$-phase difference 
between two superconducting leads.
We stress that the Josephson current of a topological and trivial Josephson junction has different periodicity with respect to the superconducting phase difference, i.e., $4 \pi$ periodicity for topological vs $2\pi$ periodicity for trivial junctions. This difference in the periodicity can be used as an experimental signature for the topological superconductivity.
Importantly, one cannot differentiate between topological and trivial Josephson junction from the amplitude of the Josephson current.

Of particular interest here is to see the extent to which 
a proximitized junction (which has a vanishing order parameter)
can, nevertheless, exhibit Josephson signatures.  
The behavior for the two different platforms is not dramatically
different. Figures~\ref{fig:4a}(a) and~\ref{fig:4b}(a) present
a plot of the energy dispersion as a function
of $k_x$ for the first and second platforms, respectively.
There are two notable effects as compared to junctions when
both superconductors have the same phase: the excitation gap
is significantly reduced and the flat bands at the edge are now
four-fold degenerate. This four-fold degeneracy 
(associated with the low-energy states outlined in red) 
appears when
the two bands just above and below the flat bands [these are
shown at the edges of the spectrum in Figs.~\ref{fig:2}(a) and~\ref{fig:3}(a)] 
move down to align with the two Majorana flat bands. This occurs precisely
when the phase difference reaches $\pi$.
There are many more subgap states in Fig.~\ref{fig:4a} than Fig.~\ref{fig:4b}
because in platform 2 the insulating region is more extended 
thereby leading to denser low-energy states.

It is useful to label the energies of these two near-by
bands as $\pm E(\phi)$ and follow their behavior as the phase
difference continuously varies. Note that these two have different
fermion parity. When $\phi$ is less than $\pi$ the band
associated with $+$ is above that with $-$. For angles between
$\pi$ and $3 \pi$, the two bands exchange places, with the $-$ band having
higher energy that the $+$ band. At $3 \pi$ the two bands will
cross again at zero energy. 

Using $E(\phi)$ we are able
to predict the behavior of 
the Josephson current
~\cite{lutchyn10,oreg10,Berg4pi,4pi}.
It follows that (unless there is a fermion-parity switch), the Josephson
current will not return to its $\phi = 0 $ value until $\phi = 4 \pi$.
This $4 \pi$ periodicity is
a well known feature of
such junctions 
~\cite{kitaev2001unpaired,Fu-2009-Josephson}
and a signature of topological order. 
What is new here is these effects occur in
a medium which has no intrinsic pairing. They are occuring strictly
via proximity coupling.

Indeed, this proximity coupling is illustrated in Figs.~\ref{fig:4a}(b)
and \ref{fig:4b}(b) which plot the position-dependent pair amplitude
for different values of $y$.
This shows the expected sign change as one crosses from one superconductor
to another.
To accomodate this overall sign switch there appear to be distinct nodal points. 
Finally, Figs.~\ref{fig:4a}(c) and \ref{fig:4b}(c) present contour plots
of the wavefunction
amplitudes associated with 
the four-fold degenerate flat bands at fixed $k_x = 0.93k_F$ and $k_x=0.96k_F$,
respectively. 
The two spots on the
left and right are the expected Majorana bound states (MBSs) coming from the
far edges of Ho, while the two in
the middle correspond to the localized wavefunctions associated with
the middle two edges of the Ho substrate. These are sometimes
described as hybridized Majorana modes~\cite{Yacoby}. 
Indeed, one can view the $4\pi$ periodicity discussed above as 
arising from these hybridized modes.

What is particularly interesting in the first platform configuration
is that even though here there is no natural junction in the middle
of the Ho substrate,
the wavefunction 
nevertheless exhibits a break into two separate contributions, 
as in platform 2.

\vskip100mm

\section{Assessing Standard Proximity Models: One dimensional junctions}

\begin{figure*}
\includegraphics[width= 6.4in]
{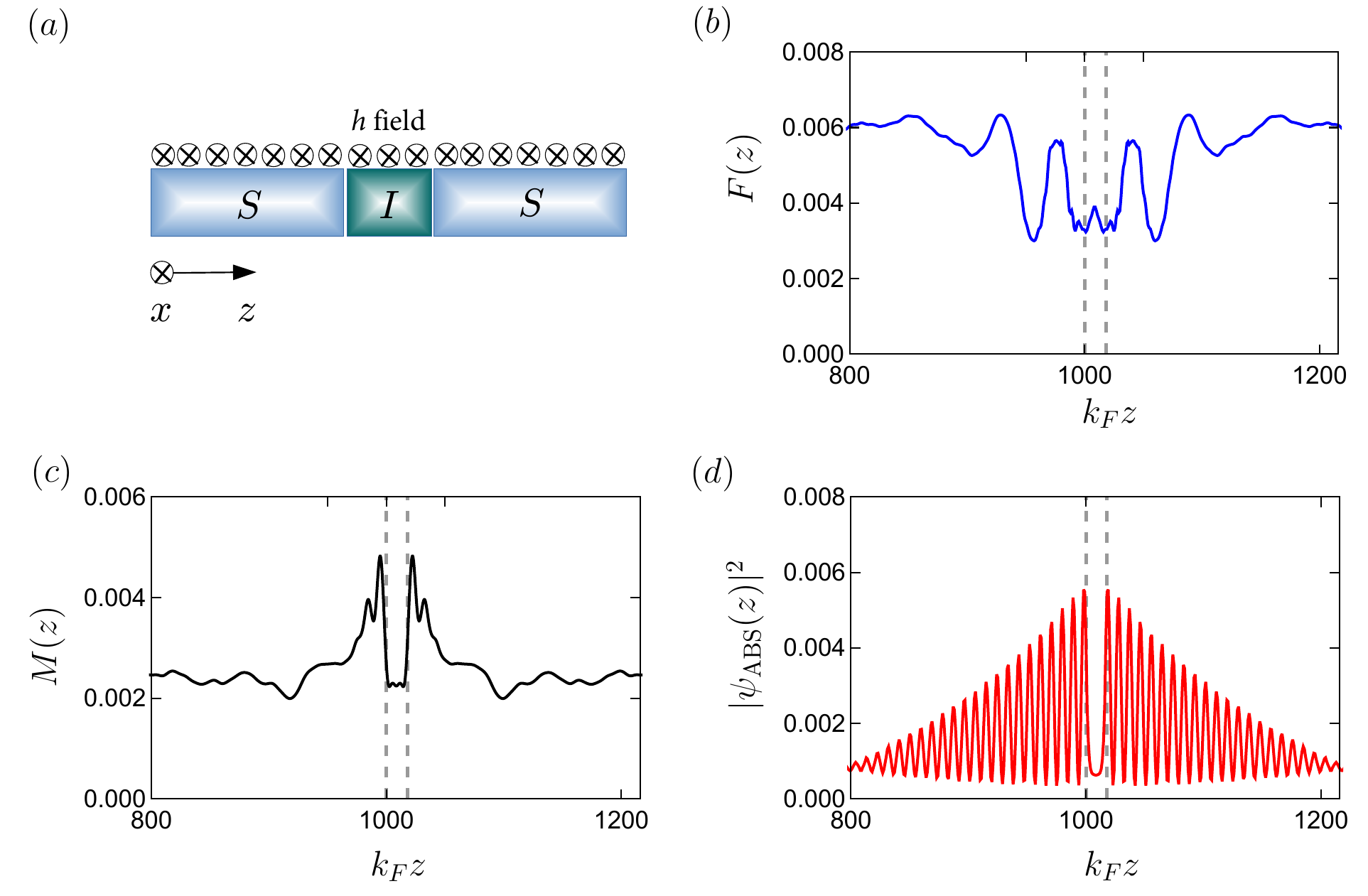}
\caption{(a) Schematic arrangement of
the effective low energy proximity model corresponding to
Platform 1, 
(b) pair amplitude profile $F(z)$, (c) profile of total magnetization $M(z)$ and (d) the zero-energy bound state wave function 
which roughly correlates with the induced magnetization. 
The dashed lines in panels (b)-(d) mark the boundaries between the superconducting and insulating regions. This figure can be compared with Fig.~\ref{fig:4d}(a).
The parameters used in the above plots are $t_\perp = 1.5$, $\ell_S = 157\xi$, $\ell_I = 2.5\xi$, $\Delta = 0.1$, $\phi = 0$, $\mu_S = 1$, $\mu_I = -1$, $\lambda_S = 0.2$, $\lambda_I = 0.2$, $h_S = 0.6$, and $h_I = 0.6$.}
\label{fig:magnetizationtopo}
\end{figure*}

\begin{figure*}
\includegraphics[width= 6.4in]
{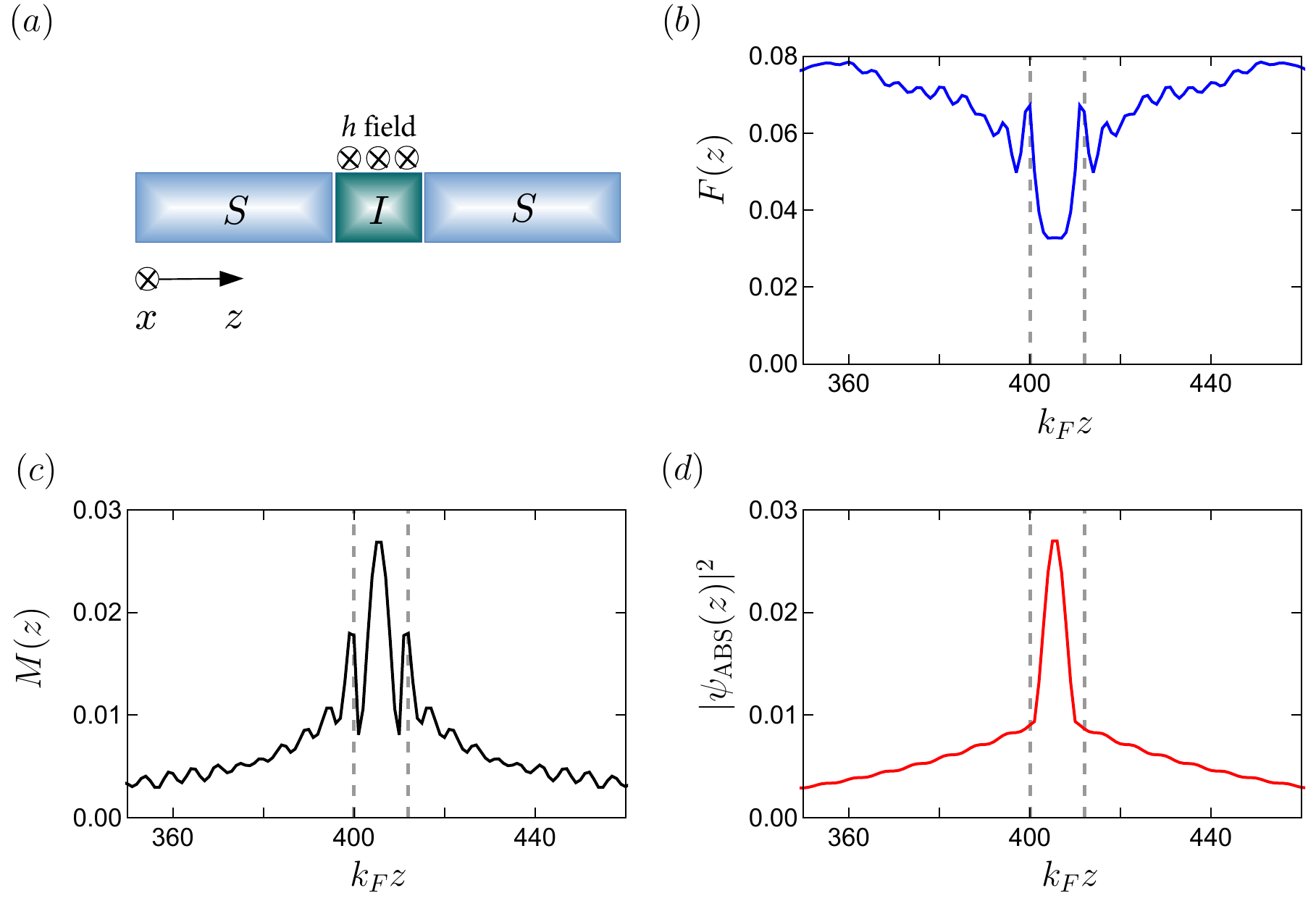}
\caption{(a) Schematic arrangement of a 1D ``trivial phase"
junction with magnetic field only in the insulating region.
This case is not related to either of the two platforms in
Figure 1, but is presented to illustrate how the bound state
wavefunctions are localized to regions with non-zero magnetic field.
(b) Pair amplitude profile $F(z)$, (c) Profile of total magnetization $M(z)$ and (d) zero-energy bound state wave function
which has an amplitude  maximum where the magnetization is maximal.
This case is non-topological, as there are no magnetic fields in S.
The dashed lines mark the boundaries between the superconducting and insulating regions. The parameters used in the above plots are $t_\perp = 1.5$, $\ell_S = 60\xi$, $\ell_I = 1.57\xi$, $\Delta = 0.1$, $\phi = 0$, $\mu_S = 1$, $\mu_I = -1$, $\lambda_S = 0.5$, $\lambda_I = 0.5$, $h_S = 0$, and $h_I = 0.6$}
\label{fig:magnetizationtrivial}
\end{figure*}

A deep understanding of the nature of proximity-induced superconductivity in systems
with combined spin-orbit coupling and Zeeman fields is
central to arriving at topological superconductors.
Rather than introducing the source of proximitization directly
as we do here, one
usually ignores the multiple layers of Figure 1, and considers an
effective SIS system with an assumed gap parameter in S.
In a nice series of papers, Stanescu and co-workers 
\cite{Saurobustness,Stanescu2,Stanescu3,Stanescu_2011} developed this approach.
They showed how to derive  
an effective low energy model
in which the superconducting degrees of freedom can be integrated
out and replaced by an interface self energy.  

Implementing their approach for the 2 different platforms
leads to 2 different quasi-one
dimensional models which (unfortunately) also need to be 
addressed numerically.
Important here is to make these simpler models compatible with
our Holmium studies by choosing
a one-dimensional spin-orbit coupling. This can be shown to
host Majorana flat bands in the topological regime.
Moreover, as shown below, we
find multiple parity switches associated with ZECs in the SIS spectrum.

Notably, these always require a sufficiently strong magnetic
field to be present inside the insulator. We, thus, presume throughout this
section that there is a non-zero field in the insulating region. 
It can be noted that this appears to be (at least) a superficial
difference between these
effective low energy models and the platform configurations
discussed in the previous section,
where no magnetic field is present in the actual insulating component of the
junction; moreover, this region
is relatively free of any magnetization [as can be seen in
Figs.~\ref{fig:2}(c) and \ref{fig:3}(c)]. Similarly 
the insulator 
does not host the zero-energy bound states in contrast to what is found
in the effective low energy models.
Rather the Holmium substrate is the active component in the junction
(just below the insulator) hosting both the magnetization and the bound state.

In the following, we focus on the effective low energy model which involves
1D spin-orbit-coupled superconducting wires~\cite{lutchyn10,oreg10} coupled infinitely along one direction (which we take to be the $x$-direction)~\cite{Wang2014}. Here we consider the system
to be infinitely thin along the $y$-direction as shown in Fig.~\ref{fig:magnetizationtopo}(a).
Similar SIS and SNS junctions have been studied in the literature
\cite{Spanish} 
which argue that
for sufficiently large Zeeman fields,
parity crossings are made possible by the nontrivial topology
in the underlying effective $p$-wave superconductor.
Here 
we find that these ZECs can also arise in the absence of SOC, hence they are not associated with topological phases.

\subsection{Effective low-energy models for including proximity}

We next study the effective low-energy approximation to
the full-proximitized SIS junction introduced in the previous section, namely an array of 1D finite-length (along $z$-direction) spin-orbit-coupled superconducting wires~\cite{lutchyn10,oreg10} which are coupled along the infinite $x$-direction~\cite{Wang2014}. 
This effective low energy model is obtained by removing the two
superconductors above the Holmium. Following the standard procedure 
for addressing proximity effects~\cite{Saurobustness, Stanescu2,Stanescu3,Stanescu_2011}, we integrate out the upper (SIS) layer  
which results in a contribution of a surface self energy in the Hamiltonian of Holmium. The self-energy is given by
\begin{equation}\label{eq:selfenergy}
\Sigma(\omega) = -|\widetilde{t}|^2 \nu\left(E_F(z)\right) \left[\frac{\omega \tau_0 + \Delta_0(z) \sigma_y\tau_y}{\sqrt{\Delta_0(z)^2-\omega^2}} + \frac{\zeta}{1-\zeta^2}\tau_z\right],
\end{equation}
where $\widetilde{t}$ is the tunneling coupling between the SIS layer and Holmium, $\Delta_0$ is the parent superconductivity, $\nu(E_F) = 2\sqrt{1-\zeta^2/\Lambda}$ is the density of states at the Fermi energy $E_F$ with $\zeta = [\Lambda - E_F(z)]/\Lambda$ and $\Lambda$ being half of the bandwidth. The second term in Eq.~\eqref{eq:selfenergy} gives rise to the proximity-induced superconductivity and the last term introduces a shift in the chemical potential of the substrate. In this way we have an effective low-energy model for the Holmium component where the region below the host superconductor 
is a proximitized superconductor while that below the insulating barrier is
treated as an insulator in a magnetic field, as shown in Fig.~\ref{fig:magnetizationtopo}(a).  This self-energy can be thought of as a consequence of the penetration of the wave function from the upper layer SIS part into the Holmium substrate. Since we are only interested in the zero-energy bound states which are the low-energy properties of the system, we can approximate $\Sigma(\omega) \approx \Sigma(0)$.

The Hamiltonian of the effective low-energy model of the Holmium is then given by
\begin{align}\label{eq:HNW}
\mathcal{H}_{1D} = &\int dz \left[\left(-\frac{\partial_z^2}{2m} - \mu(z)\right)+H_{\mathrm{\parallel}}\right. \nonumber\\
&\hspace{1 cm}\left.+\left(\Delta(z)\psi^{\dagger}_{\uparrow}(z)\psi_{\downarrow}^\dagger(z) + \mathrm{H.c.}\right)\right],
\end{align}
where $\psi_{\sigma}(z) [\psi^\dagger_{\sigma}(z)]$ is the annihilation (creation) operator of an electron at position $z$ with spin $\sigma = \uparrow, \downarrow$ and $\Delta(z) = |\widetilde{t}|^2\nu(E_F)\theta(\Delta_0(z))$ is a proximity-induced $s$-wave pairing potential. 

Note that the chemical potential $\mu(z)$ and pairing potential $\Delta(z)$ vary along the $z$ direction [see Fig.~\ref{fig:magnetizationtopo}(a)] where $\mu_I < 0$, $\mu_S > 0$, $\Delta_S > 0$, and $\Delta_I = 0$. In the above, we have used the subscripts I and S to denote the quantities corresponding to the I and S regions, respectively in
Fig.~\ref{fig:magnetizationtopo}(a).
  
 The term 
\begin{align}
H_{\mathrm{\parallel}} = &-i\lambda \left(\psi^\dagger_{\uparrow}(z)\psi_{\uparrow}(z+\hat{z})- \psi^\dagger_{\downarrow}(z)\psi_{\downarrow}(z+\hat{z})\right) + \mathrm{H.c.} \nonumber\\
& + h \left(\psi^\dagger_{\uparrow}(z) \psi_{\downarrow}(z) + \psi^\dagger_{\downarrow}(z) \psi_{\uparrow}(z)\right),  
\end{align}
contains the spin-orbit ($\lambda$) and Zeeman ($h$) coupling. Note that this nanowire Hamiltonian is equivalent to the proximity-induced superconducting ferromagnet at a conical opening angle $\alpha = \pi/2$~\cite{ourPRB} as discussed in the previous section. 

We introduce a coupling in the array of the 1D wires along the $x$-direction via the term
\begin{equation}
H_{\perp} = -t_\perp \int dx \left[\psi^\dagger(x,z) \psi(x+\hat{x},z)+ \mathrm{H.c.} \right].
\end{equation}
Here we have defined $\hat{x}$ and $\hat{z}$ to be the unit vectors along the $x$ and $z$ directions, respectively.
Since the Hamiltonian is translationally invariant along the $x$-direction, the Hamiltonian of the above 2D system can be dimensionally reduced into a sum of 1D wires each with specific values of $k_x$, i.e.,
\begin{align}\label{eq:H2D}
\mathcal{H}_{\mathrm{2D}} = \int dk_x \mathcal{H}_{\mathrm{1D}} (k_x).
\end{align} 
where $\mathcal{H}_{1D}(k_x)$ is the same as Eq.~\eqref{eq:HNW} written in momentum space with the chemical potential replaced by $\mu \rightarrow \widetilde{\mu}(k_x)= \mu +2t_\perp \cos(k_x)$. When the magnetic field is tuned above its critical value $h > |\Delta|$, the superconducting region of the array of 1D wires can support any integer $Z$ gapless Majorana zero modes thus it belongs to the topological class BDI~\cite{schnyder08, kitaev09,Tewari12}. For $h < |\Delta|$, the system is in the trivial phase and gapped. The gap closes when $h> |\Delta|$ where the system becomes topological with Majorana flat bands in the $E$ versus $k_x$ spectrum. These Majorana flat bands can be found in the $k_x$ region where $h^2 > \widetilde{\mu}(k_x)^2 + \Delta^2$.  We note that the gap closes at the $k_x$ values where $\widetilde{\mu}(k_x)^2 + \Delta^2 = h^2$. For the case where $ |\sqrt{h^2 - \Delta^2}-|\mu| | \leq 2t_\perp \leq (|\mu| + \sqrt{h^2 - \Delta^2})$,  there are two gap closing points and for the case where $|\mu| + \sqrt{h^2-\Delta^2}  \leq 2t_\perp$, there are four.

\subsection{Numerical Results: Effective low-energy models}
We focus on numerical solutions of the BdG equations given in Eqs.~\eqref{eq:HNW}-\eqref{eq:H2D}. In particular, we compute the zero-energy bound state wavefunction, proximitized gap [Eq.~\eqref{eq:Deltaself}],  and magnetization [Eq.~\eqref{eq:magnetization}]. Importantly, here, too, we find ZECs in the spectrum which signify QPTs.

In Figs.~\ref{fig:magnetizationtopo}(a)-(d)
we present solutions of
the effective low energy model corresponding to Platform 1. 
Here, the insulating  region has a constant negative chemical potential 
$\mu_I < 0$ with a length $\ell_I\sim \xi$ where $\xi = 2/(\pi \Delta)$ is the superconducting coherence length with $\Delta$ being the superconducting gap.
Multiple zero energy bound states of the effective model are found to appear
once
the Zeeman field in the insulating region of an SIS junction is increased to
the critical value ($h_I = |\mu_I|$), where $\mu_I$ is
the chemical potential in I.

Figure~\ref{fig:magnetizationtopo}(a) shows the junction configuration where
the magnetic field is naturally present in the 
insulating region.
The pairing potentials on the left and right  superconducting regions are taken to be $\Delta$ and $\Delta e^{i\phi}$, respectively. For simplicity, the results here are presented for
the case $\phi =0$.

To relate to our more complete proximity calculations,
we plot 
the associated pairing amplitudes 
$F({\bf r})$,
[see Fig.~\ref{fig:magnetizationtopo}(b)], the magnetization
[see Fig.~\ref{fig:magnetizationtopo}(c)] and (for one particular
bound state) the zero-energy wavefunction amplitude
as functions of position throughout
the junction [see ~\ref{fig:magnetizationtopo}(d)].
This figure can then be compared with Fig.~\ref{fig:4d}(a).
Just as in Fig.~\ref{fig:4d}(a), the magnetization and the wavefunction 
amplitude are correlated with each other: they assume their
maximum values at the same position. One can compare with
Fig.~\ref{fig:4d}(a) where the counterparts from the full
proximity calculation exhibit 
a maximum in the junction center, whereas in Fig.~\ref{fig:magnetizationtopo}, they
both have minima. The pairing amplitudes shown in Fig.~\ref{fig:4d}(a)
and in Fig.~\ref{fig:magnetizationtopo}(b) are more obviously similar, as they both exhibit
a dip in the center of the junction.
It should be noted, as can be seen from the plot, the pairing penetrates into
the insulator.

If the magnetic field is absent in the I region of the effective
low energy model, then the non-topological zero energy bound states are no longer present.
This case, which would be associated with platform 2, does not
yield a discrete zero energy crossing in the dispersion as in
the full proximity case of
Fig.~\ref{fig:2}(a). 
Rather it is associated with a Majorana flat band.
In this way, we infer that these effective low energy models do not
always accomodate the same detailed physics as in a more realistic proximity
junctions.

Finally, we present results
in Fig.~\ref{fig:magnetizationtrivial}
for a simple case of a trivial SIS junction where the superconductor
contains SOC but no magnetic field. For this configuration
the magnetic field
is restricted to be inside the insulating region. This case
is unrelated to the 2 holmium platforms considered throughout the
paper. Nevertheless, it serves to illustrate the importance of
a non-zero Zeeman field in hosting non-topological quantum phase transitions.
We have seen in the second platform configuration
(Figures \ref{fig:3} and \ref{fig:4d}(b))
that the bound state wave function is essentially excluded from
regions in the junction where the field vanishes. Thus the
wave function does not overlap the insulating region.
We see in
Figure \ref{fig:magnetizationtrivial} a similar effect. Here,
again following the behavior of the magnetic field, the
bound state wave function is essentially confined to the insulating
region and excluded from the host superconductors where the field
vanishes.

We note that this case is closer to that studied in
Refs.~\cite{LiuSau,Spanish}.
What is particularly intriguing about this situation is that
it can be thought of as a ``quantum dot" system where
Coulomb blockade effects are absent. Generally the presence
of Coulomb
blockade physics is used \cite{Glazman2} to argue that
Kondo physics is driving the QPT in quantum dots. This
follows by using a Schrieffer-Wolff transformation to
convert the dot Hamiltonian to a magnetic impurity model.
What is shown here and in Ref. \cite{LiuSau} is that
Coulomb effects are not essential for arriving at these quantum
phase transitions
in quantum dots. 

The junction configuration of interest is shown in
Figure \ref{fig:magnetizationtrivial}(a) whereas
Figure \ref{fig:magnetizationtrivial}(b) presents a plot of the self-consistent
pair amplitude calculated using Eq.~\eqref{eq:Deltaself}. As can be seen from the plot, the gap penetrates into
the insulator. Figure \ref{fig:magnetizationtrivial}(c) shows the magnetic screening cloud
or magnetization $M(z)$ throughout the junction with its components calculated using Eq.~\eqref{eq:magnetization}. The
magnetization is largest in the insulating region as a consequence of
the magnetic field there.
Finally in
Figure \ref{fig:magnetizationtrivial}(d)
we present a plot of the amplitude of the
wavefunction corresponding
to a prototypical zero-energy bound [$\psi_{\mathrm{ABS}}(z)$]. We see
that the magnetization and
the wavefunction are spatially correlated and peaked in the insulating
regime, reflecting the presence of the magnetic field which only
appears in this region.

\section{Conclusions}

\subsection{Comparison of the full proximity results with the effective
model}

It is interesting to focus on the behavior in the vicinity of the
various interfaces in the complex Josephson junctions we
consider. Note that the S-Ho interface
is effectively absent in the low energy models used for addressing proximity
coupling (see Section IV, where the contribution from S has been integrated out.)
On the otherhand, it is accessible in the full proximity
calculations and both Figs.~\ref{fig:2} and \ref{fig:3} show that the induced magnetization
barely penetrates into the superconducting region. This 
is due to the fact that the exchange interaction is local and present only in the Ho layer~\cite{Bergeret}.
Rather the magnetization is confined to Ho. In platform 2 it
is locally depressed
in the insulating region, where Ho is completely absent; it, nevertheless, recovers to the full
bulk value associated with Ho at the sample ends far from the insulator.
By contrast in platform 1 the magnetization is increased in
the vicinity of (but below) the insulator, acting much as a local magnetic moment.

The pair amplitude undergoes a more non-monotonic behavior associated with the
S-Ho interface which
is missing in the low energy approximate models. These oscillations
are well known (see, for example~\cite{halterman2}). In this way there is a depression in the pairing
amplitude very close to the S-Ho interface, but it recovers to become
rather strong somewhat below.
Importantly, the wavefunctions for the QPT in Platform 1 are localized in Ho in the regime
where the pairing amplitude is maximal.

At some level there is consistency between the effective models and the
full proximity calculations, as we find in platform 1 (through both approaches)
that there are non-topological zero energy crossings. Similarly we find
in platform 2 (through both approaches) that the crossings there are
only topological.  What is important to stress, however, is that the full
proximity models are more complete because they self consistently
establish the degree and even the presence of proximitization. In the effective models one
usually introduces a phenomenological pairing gap parameter $\Delta$
as in Eq.~(\ref{eq:HNW}), whose size can only be obtained from the
full proximity calculations. We emphasize this size is critical in determining
the conditions for topological and trivial phases. 
These differences are also made clear by contrasting
Figs.~\ref{fig:4d}(a) and \ref{fig:magnetizationtopo}.

\subsection{Physical Picture of the quantum phase transition}

By way of summary, it is useful to revisit the question of
what is the physical mechanism responsible for these
non-topological zero energy bound states found in either the full proximity Josephson case
or with the effective low energy model. We have shown throughout that
there is a clear correlation between the inhomogeneous magnetization
and the amplitude of the bound state wavefunction.
We associate a proximity-induced
``magnetic defect" with the insulating region of the junction.
In the presence of magnetic fields, the insulator effects
a local reorganization of the magnetization in the proximitized medium.
We emphasize
that our insulating barriers are \textit{non-magnetic}. The
proximity-induced ``magnetic" defects which we refer to
as magnetization inhomogeneities, have a different origin from
the well-studied magnetic impurities associated with the Shiba
scenario \cite{shiba1968classical}.

Nevertheless
these Shiba or ``external" impurities provide a useful template
for understanding zero-energy crossings. These crossings are found to occur
by tuning the strength of
the effective impurity exchange interaction~\cite{yu1965bound,shiba1968classical,Sakurai}.
Using this template in the present situation,
it is the magnetic field (in units
of the insulating chemical potential) which 
provides a mechanism for the energy level crossings and
parity shifts in our proximitized Josephson junctions. 
At a critical magnetic field
the energies of the superconducting junction
states with $n$ and $n+1$ electrons cross.
The parity switch of the crossing
reflects the fact that in a
region of rapidly varying magnetization it may be energetically more favorable to
invert the relative order of different states (having different parities).
This is
implemented by adding an
additional fermion spin or single particle excitation, associated with one less
pair and one extra spin.

It is useful to contrast our effective low energy proximity model results with 
the case of a delta-function non-magnetic impurity studied in Ref.~\cite{Sau-Demler}. The present picture applies to a finite-length insulating 
barrier and we find that this zero-energy crossing  can appear as long as the magnetic field in the insulating region is sufficiently large ($h_I > |\mu_I|$).The crossings we find do not require the superconducting region to be topological, as in Ref.~\cite{Sau-Demler}.
Similarly, there are studies of multiple magnetic impurities, or
impurity chains in the literature \cite{Kristofer2017}. While
we also consider magnetization defects of extended size, ours
are not externally inserted but are proximity-induced
through full self consistency. They arise because of the presence
of magnetic fields in (proximitized) superconductors, which, 
around an insulating barrier, serve to induce an inhomogeneous
magnetization. 

Finally we note that there is a literature which is closer to
the issues presented in the present paper. This deals with
Andreev bound states in the presence of magnetic
fields \cite{LiuSau,Spanish}.
The (necessarily) numerical findings from this 
body of work
show how these bound states
appear as functions of junction magnetic field and chemical potentials
but without establishing detailed microscopic governing equations.
Indeed, it has been argued both theoretically \cite{Spanish} 
and experimentally \cite{Lieber2017} that there are similarities
between Andreev bound states
and those associated with magnetic impurities. A 
notable contribution
from the present paper was to present a ``missing link"
which explains the similarity. We did this
by identifying the role of the proximity-induced
magnetization
$\bf{m}(\bf{r})$ arising in an Andreev configuration, 
which is something which was
previously of interest only to the superconducting spintronics
community.

\subsection{Summary}

In this paper we addressed fairly realistic proximitized Josephson junctions
which contain the necessary features (both Zeeman and spin-orbit coupling)
to produce
topological superconducting phases.
While we have focused on a particular example involving the conical
magnet Ho, we expect our findings to be more general.
These junctions contain multiple
non-topological zero-energy bound states associated with fermion-parity switches in
quantum phase transitions.
They are particularly important because
they have the potential to lead to ``false positives"
in reports for Majorana bound states. Thus, understanding their
origin more microscopically provides a central motivation
for our work here.

To understand these quantum phase transitions,
we have presented a full proximity treatment of  multi-component
Josephson junctions. Despite the many papers concerned with
topological junctions, a 
detailed and precise proximity analysis appears to be otherwise lacking.
It, moreover, provides a valuable check on widely used
effective low energy models which we examine here.
We show how it is useful to consider self-consistently derived 
quantities from
the BdG analysis such as the
screening cloud magnetization 
$\bf{m}(\bf{r})$,
and induced pairing amplitudes
$F({\bf r})$.
These allow us 
to understand the appearance of zero-energy
bound states and to correlate their areas
of confinement to the
spatial dependences of these properties.
In this way, we
arrive at a generalization of
the magnetic impurity scenario for these quantum phase transitions,
but here with a self consistent and proximity-induced magnetic defect.

\textit{Acknowledgements.--}
We thank Erez Berg, Shinsei Ryu, J. Robinson, Alexa Galda, Jay Sau and Michael Levin for helpful conversations. This work was supported by NSF-DMR-MRSEC 1420709. C.-T.W. is supported by the MOST Grant No. 106-2112-M-009-001-MY2C. We acknowledge the University of Chicago Research Computing Center for support of this work.

\bibliography{Review2}

\end{document}